\documentclass[twocolumn, reprint,
aip,
jap,
amsmath,amssymb
]{revtex4-1}
\usepackage{graphicx,color}
\usepackage{amsmath}
\usepackage{natbib}
\usepackage{epsfig}
\usepackage{epstopdf}
\usepackage{hyperref}

\begin{document}

\title{{\color{blue} Onset of transverse (shear) waves in strongly-coupled Yukawa fluids}}

\author{Sergey A. Khrapak}
\email{Sergey.Khrapak@dlr.de}
\affiliation{Institut f\"ur Materialphysik im Weltraum, Deutsches Zentrum f\"ur Luft- und Raumfahrt (DLR), 82234 We{\ss}ling, Germany}

\author{Alexey G. Khrapak}
\affiliation{Joint Institute for High Temperatures, Russian Academy of Sciences, 125412 Moscow, Russia}

\author{Nikita P. Kryuchkov}
\affiliation{Bauman Moscow State Technical University, 105005 Moscow, Russia}

\author{Stanislav O. Yurchenko}
\email{st.yurchenko@mail.ru}
\affiliation {Bauman Moscow State Technical University, 105005 Moscow, Russia}

\date{\today}

\begin{abstract}
A simple practical approach to describe transverse (shear) waves in strongly-coupled Yukawa
fluids is presented. Theoretical dispersion curves, based on hydrodynamic consideration, are shown to compare favorably with existing numerical results for plasma-related systems in the long-wavelength regime. The existence of a minimum wave number below which shear waves cannot propagate and its magnitude are properly accounted in the approach. The relevance of the approach beyond plasma-related Yukawa fluids is demonstrated by using experimental data on transverse excitations in liquid metals Fe, Cu, and Zn, obtained from inelastic x-ray scattering. Some potentially important relations, scalings, and quasi-universalities are discussed. The results should be interesting for a broad community in chemical physics, materials physics, physics of fluids and glassy state, complex (dusty) plasmas, and soft matter.
\end{abstract}

\maketitle

\section{Introduction}

The description of multi-component charged particle systems can be in some cases considerably simplified by using the concept of Yukawa one-component plasma (YOCP). In this concept the lighter components are usually treated as a mobile neutralizing medium, which provides  screening of the charges carried by the less mobile heavy component. The system then represents a collection of particles of charge $Q$ interacting via the repulsive pairwise interaction potential of the (Debye-H\"uckel or Yukawa) form $\phi(r)=(Q^2/r)\exp(-r/\lambda)$, where $\lambda$ is the screening length, representing the only remaining characteristic of the neutralizing medium. Equilibrium properties of such systems are conventionally described  by the two dimensionless parameters: the coupling parameter $\Gamma=Q^2/aT$ and the screening parameter $\kappa=a/\lambda$, where $T$ is the system temperature (in energy units), $n$ is the particle density, and $a=(4\pi n/3)^{-1/3}$ is the Wigner-Seitz radius.
The YOCP are often used as a simplified representation of strongly coupled plasmas, complex (dusty) plasmas, and colloidal dispersions.~\cite{Hansen2000, LowenJPCM1992, FortovUFN2004, FortovPR2005, IvlevBook}
Thermodynamics and phase portrait of idealized Yukawa systems have been extensively studied in the literature, see for example Refs.~\onlinecite{HamaguchiJCP1996, HamaguchiPRE1997,KhrapakPRE2014,ToliasPRE2014, ToliasPoP2015, KhrapakISM, KhrapakPRE2015, KhrapakJCP2015, KhrapakPoP2015,VeldhorstPoP2015,YurchenkoJCP2014, YurchenkoJCP2015, YurchenkoJPCM2016, KhrapakPoP2017_Grun} and references therein.

One of the important topics in investigating strongly-coupled systems is related to collective dynamics and collective modes.~\cite{DonkoJPCM2008,MerlinoJPP2014,ClerouinPRL2016,ArkhipovPRL2017,KhrapakPRE2018,KhrapakJCP2018}
It is well known that a dense fluid, not too far from the fluid-solid phase transition, can sustain one longitudinal and two transverse modes.
The existence of transverse modes in fluids is a consequence of the fact that its response to high-frequency short-wavelength perturbations is similar to that of a solid body.~\cite{ZwanzigJCP1965}
The main difference from transverse waves in solids is the existence of a minimum (critical) wave number $q_*$, below which shear waves cannot propagate (here and below $q=ka$ denotes the reduced wave number).
This phenomenon, referred to as the $q$-gap in the transverse mode, is a well known property of the fluid state.~\cite{HansenBook,BolmatovPCL2015,Trachenko2015,YangPRL2017}
It applies to conventional neutral fluids, but also to the charged plasma-related systems at strong coupling.
Not surprisingly, it received considerable attention in the context of complex (dusty) plasmas within the YOCP concept.~\cite{MurilloPRL2000,OhtaPRL2000,HamaguchiPS2001,HouPRE2009,GoreePRE2012}

The purpose of this article is to present simple yet accurate analytical approach to describe quantitatively the transverse dispersion relation in strongly coupled YOCP. Recent developments related to the thermodynamics, transport and collective modes in Yukawa fluids are combined in order to derive useful practical expressions describing the transverse mode, including the existence and location of the critical wave number $q_*$.
Extensive comparison with results from different numerical simulations demonstrates the adequacy of the proposed approach. It is shown that the approach is particularly well suitable at strong coupling (where the $q$-gap is relatively narrow) and describes very well the long-wavelength portion of the dispersion relation.
Moreover, some of the scalings and tendencies  resulting from the approach, are likely relevant beyond the plasma-related context.
This is demonstrated by applying the approach to describe transverse oscillations of liquid Fe, Cu, and Zn, recently measured experimentally using inelastic x-ray scattering.\cite{Hosokawa2015}
Towards the end of the paper we discuss several important points related to the correct interpretation of the transverse spectra.

\section{Description of the transverse mode}\label{Theo}

A very simple and convenient for practical applications expression for the dispersion relation of transverse waves in strongly coupled Yukawa fluids has been recently proposed.~\cite{KhrapakPoP2016} It reads
\begin{equation}\label{T1}
\omega_{\rm QCA}^2(q)=\omega_{\rm p}^2e^{-R\kappa}\left(1+R\kappa\right)\left(\frac{1}{3}+\frac{\cos Rq}{R^2q^2}-\frac{\sin Rq}{R^3q^3} \right),
\end{equation}
where $\omega_{\rm p}=\sqrt{4\pi Q^2 n/m}$ is the plasma frequency scale, $m$ is the particle mass, and $R$ is the so-called correlational hole radius, expressed in units of $a$. At strong coupling this radius depends only on the screening parameter and is approximately given by~\cite{KhrapakAIPAdv2017}
\begin{equation}\label{R}
R(\kappa)\simeq 1+\frac{1}{\kappa}\ln \left[\frac{3 \cosh (\kappa)}{\kappa^2}-\frac{3 \sinh
(\kappa)}{\kappa^3}\right].
\end{equation}
For weak screening, further simplification is possible and yields $R(\kappa)\simeq 1+\kappa/10$.

These expressions follow from the combination of the quasi-crystalline approximation (QCA),~\cite{Hubbard1969} known as the quasi-localized charge approximation (QLCA) in the plasma-related context,~\cite{GoldenPoP2000} with the simplest step-wise model of the radial distribution function (RDF) of the form $g(x)=\theta(x-R)$, where $x=r/a$. The main idea behind this simplification is that since the function $g(r)$ appears under the integral, an appropriate model for $g(r)$ can be constructed. The main requirement is to correctly reproduce the integral properties, but not to describe very accurately the actual structural properties of the system. The simplest trial RDF is clearly of the form specified above. This approach has been shown to be rather useful in describing long-wavelength dispersion relations in various systems with sufficiently soft interactions in both three dimensions (3D)  and two dimensions (2D).~\cite{KhrapakAIPAdv2017,KhrapakPoP2016,2DOCP,KhrapakPRE2018,KhrapakJCP2018}

Equation (\ref{T1}) does not include the term responsible for the kinetic effects [$(q/a)^2v_{\rm T}^2$, where $v_{\rm T}=\sqrt{T/m}$ is the thermal velocity scale], which is numerically small in the strongly coupled regime (where transverse waves can propagate).

Another of the well known weaknesses of the QCA approach is that it cannot account for the disappearance of the transverse mode at long-wavelengths and the existence of the $q$-gap. Below we discuss a simple practical approach to overcome this deficiency, which is based on the ``hydrodynamic'' account of the $q$-gap using independent data for shear viscosity.

It is conventional to get more insight about the dispersion of collective modes from the analysis of current autocorrelation functions.~\cite{BalucaniBook} Within the framework of the generalized hydrodynamics, supplemented by a single exponential memory function approximation, the transverse current spectrum is~\cite{BalucaniBook,HansenBook,AkcasuPRA1970,UpadhyayaNJP2010,MithenPRE2014}
\begin{equation}
C_t(q,\omega)\propto \frac{\tau_{\rm R}\omega_t^2(q)}{\omega^2+\tau_{\rm R}^2[\omega^2-\omega_t(q)^2]^2},
\end{equation}
where $\omega_t^2(q)$ is the normalized second frequency moment of the transverse current spectrum, which can be expressed in terms of $\phi (r)$ and $g(r)$.~\cite{BalucaniBook} It is also related to the {\it generalized q-dependent} high-frequency (instantaneous) shear modulus via $(q/a)^2G(q)/mn = \omega_t^2(q)$.\cite{BalucaniBook,Schofield1966,Nossal1968} 
In fact, the configurational part of $\omega_t(q)$ coincides with the dispersion relation derived within the QCA (QLCA) approach.  
Thus, as long as the kinetic term can be neglected, we can use $\omega_t^2(q)=\omega_{\rm QCA}^2(q)$. This simplify considerably calculations and we use this condition henceforth. The relaxation time $\tau_{\rm R}$ appears in the approach as a $q$-dependent Maxwell relaxation time,
\begin{equation}\label{tauR}
\tau_{\rm R}(q)= \frac{\eta(q)}{G(q)},
\end{equation}
where $\eta(q)$ is the $q$-dependent coefficient of shear viscosity. Differentiating $C_t(q,\omega)$ with respect to $\omega$, it is easy to demonstrate that $C_t(q,\omega)$ has a peak at non-zero frequency, provided  $\omega_{\rm QCA}^2(q)\tau_{\rm R}^2(q)>1/2.$ If it exists, the peak is located at
\begin{equation}\label{T2}
\omega_{\max}^2(q)=\omega_{\rm QCA}^2(q)-\frac{1}{2\tau_{\rm R}^2(q)}.
\end{equation}
In the long-wavelength regime $q\rightarrow 0$ we have $\omega_{\rm QCA}^2(q)\simeq (q/a)^2c_t^2$ and $G(0)/mn = c_t^2$, where $c_t$ is identified as the transverse sound velocity. The relaxation time tends to its long-wavelength limit $\tau_{\rm R}(0)=\eta(0)/G(0)$, where $\eta(0)\equiv \eta$ is the static shear viscosity. The dispersion relation becomes~
\begin{equation}\label{Hyd}
\omega_{\max}\simeq \sqrt{c_t^2(q/a)^2-\frac{1}{2\tau_{\rm R}^2(0)}}.
\end{equation}
This expression often appears in the literature with the coefficient $1/4$ instead of $1/2$ in the last term.  \cite{YangPRL2017,TrachenkoPRE2017,OhtaPRL2000,KawPoP2001}

In the strongly coupled regime and not at too short wavelengths, the term  $1/2\tau_{\rm R}^2(q)$ in Eq.~(\ref{T2}) is only expected to be important near the onset of the transverse mode, because $\omega\tau_{\rm R}(q)\gg 1$ otherwise. In this regime the $q$-gap is relatively narrow. Mathematically this implies $q_*\lesssim 1$ and this allows us to neglect $q$-dependence of the relaxation time, fixing it as $\tau_{\rm R}=\eta/G(0)$. Equation (\ref{T2}) with a {\it fixed} relaxation time is to some extent similar to heuristic approaches suggested previously.~\cite{HouPRE2009,KhrapakIEEE2018} The cutoff wave number for the onset of the transverse mode in this approximation reads
\begin{equation}\label{q*}
 q_* = \frac{a}{\sqrt{2}\tau_{\rm R}c_t}=\frac{a mn c_t}{\sqrt{2}\eta}.
\end{equation}

The main purpose of this article is to verify that a simple pragmatic approximation formulated above is able to describe quantitatively the transverse waves in strongly coupled plasma fluids. This is particularly important, because different definitions of the relaxation time $\tau_{\rm R}$ can be found in the literature.~\cite{YangPRL2017,KawPoP1998,KawPoP2001,BrykPRL2018,YangPRL2018} In the rest of the paper we demonstrate by numerous comparisons with existing numerical results that our approach represents a useful practical tool to describe the long-wavelength portion of the transverse dispersion curves in plasma-related fluids. Moreover, since the approach is not heavily based on plasma-related specifics, it is reasonable to expect that it can be useful in some other situations, too. As an example, we will document the consistency of the approximation with the observed transverse modes dispersion in liquid metals.

\section{Viscosity and shear modulus}

To proceed further we need to relate the shear viscosity and shear modulus to the state variables $\kappa$ and $\Gamma$ characterizing Yukawa systems. The reduced shear viscosity $\eta_*$ is introduced via $\eta=\eta_{*}m v_{\rm T} n^{2/3}$ (this is sometimes referred to as the Rosenfeld's normalization~\cite{RosenfeldJPCM1999}).  For the latter we use a general scaling with the melting temperature proposed recently by Costigliola {\it et al.}~\cite{CostigliolaJCP2018} for dense neutral fluids.
It has been demonstrated to apply well to Yukawa fluids and has a form~\cite{YukawaViscosity}
\begin{equation}\label{visc}
\eta_*\simeq 0.126\exp\left( 3.64\sqrt{\Gamma/\Gamma_{\rm m}}\right),
\end{equation}
where $\Gamma_{\rm m}$ is the coupling parameter at the fluid-solid phase transition (the subscript ``{\rm m}'' stands for melting). The dependence $\Gamma_{\rm m}(\kappa)$ was obtained in molecular dynamics (MD) simulations~\cite{HamaguchiJCP1996,HamaguchiPRE1997} and fitted by a simple formula~\cite{VaulinaJETP2000,VaulinaPRE2002}
\begin{equation}
\Gamma_{\rm m}(\kappa)\simeq \frac{172 \exp(\alpha\kappa)}{1+\alpha\kappa+\tfrac{1}{2}\alpha^2\kappa^2}.
\end{equation}
The factor $\alpha=(4\pi/3)^{1/3}\simeq 1.612$ is the ratio of the characteristic interparticle separation $\Delta=n^{-1/3}$ to the Wigner-Seitz radius $a$. This  formula works well in the regime $\kappa\lesssim 5$, which is relevant to the most experiments with strongly-coupled complex (dusty) plasmas, as well as for approximate analysis of metals' behavior.
The transverse sound velocity (and hence the shear modulus), can be estimated from
\begin{equation}
c_t^2 = \frac{\omega_{\rm p}^2a^2}{30}R^2(1+R\kappa)\exp(-R\kappa),
\end{equation}
which follows directly from Eq.~(\ref{T1}).

\section{Results}

For plasma-related systems, relaxation time is conventionally expressed in units of inverse plasma frequency. We get
\begin{equation}\label{TR2}
\tau_{\rm R}\omega_{\rm p}\simeq 2.79 \frac{\eta_*\sqrt{\Gamma}}{(c_t/v_{\rm T})^2}.
\end{equation}
Note another useful relation: $c_t/v_{\rm T}=\sqrt{3\Gamma}(C_T/\omega_{\rm p}a)$. Now all the required quantities can be evaluated for a given Yukawa fluid state point ($\kappa$, $\Gamma$). The detailed comparison with available numerical data can be performed.

\begin{figure}[!t]
\includegraphics[width=85mm]{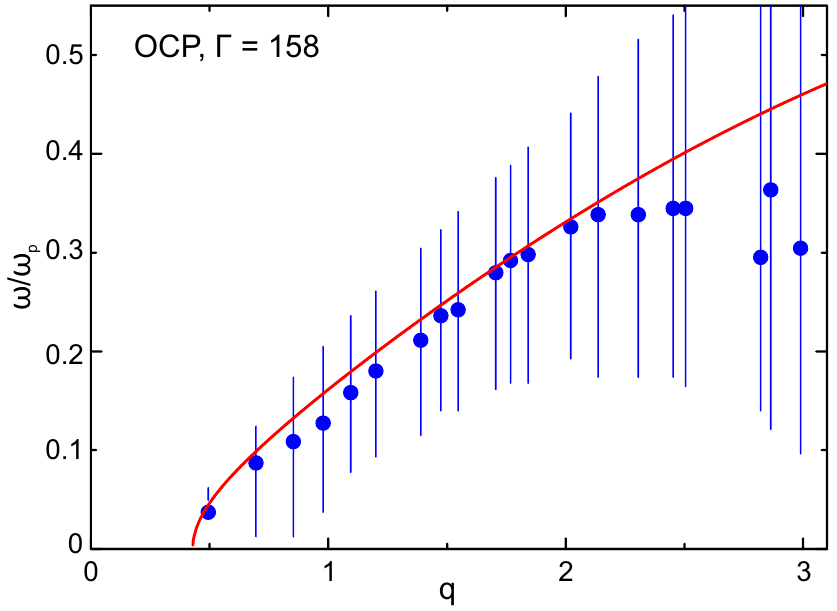}
\caption{Dispersion of the transverse mode of the strongly coupled  OCP fluid with $\Gamma= 158$.  Symbols are the results from MD simulation by Schmidt {\it et al}.~\cite{SchmidtPRE1997} The solid curve is plotted using Eq.~(\ref{T2}). In this case $\tau_{\rm R}\omega_{\rm p}=9.1$. }
\label{Fig1}
\end{figure}

We start with the one-component plasma (OCP) limit. The dispersion relation is readily obtained from (\ref{T2}) by taking the $\kappa=0$ limit in Eq.~(\ref{T1}). Figure~\ref{Fig1} shows the comparison of the resulting theoretical dispersion relation with the results from MD simulations by Schmidt {\it et al}.~\cite{SchmidtPRE1997} The agreement between theory and simulations is excellent for sufficiently long wavelength with $q\lesssim 2$.

\begin{figure}[!t]
\includegraphics[width=85mm]{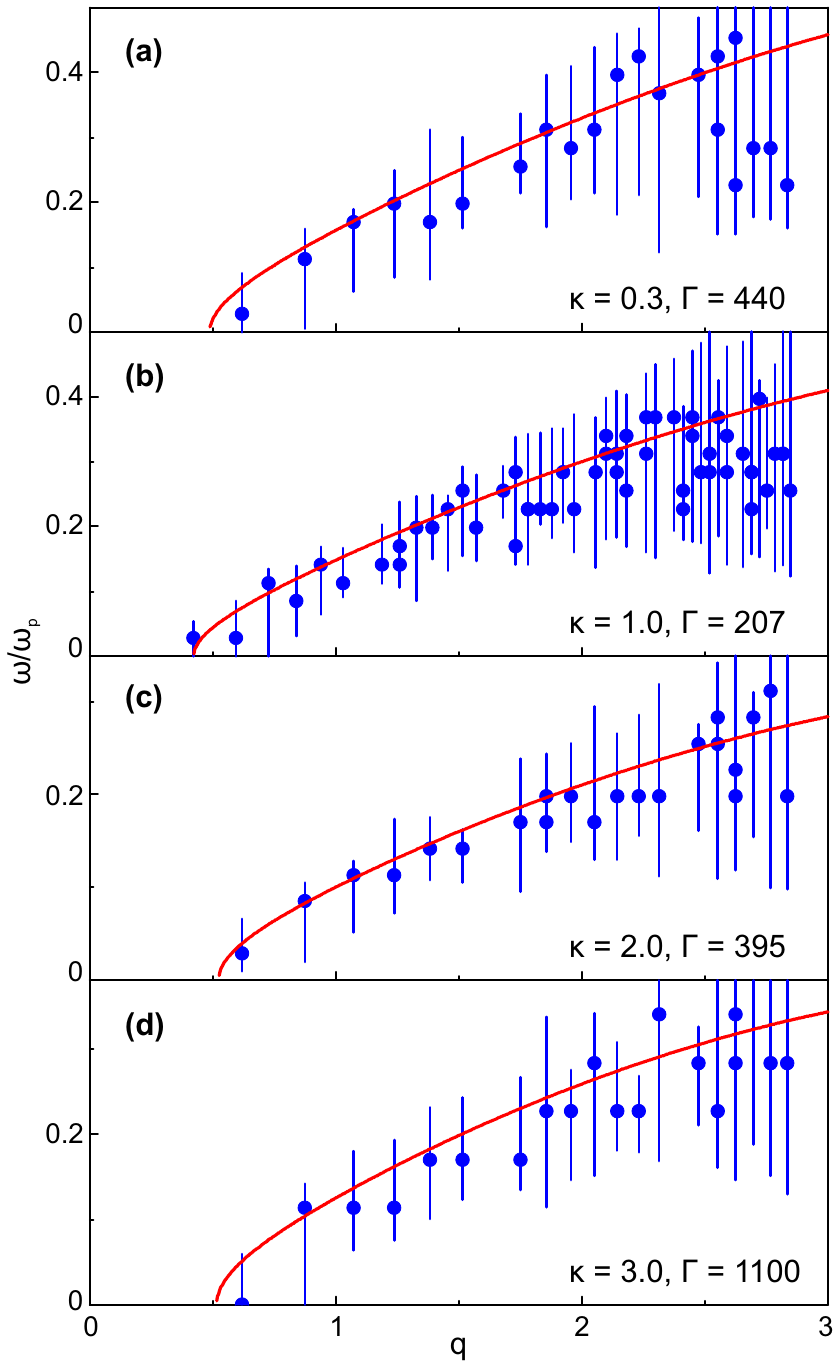}
\caption{Dispersion of the transverse waves in strongly coupled Yukawa fluids near the melting temperature (the corresponding values of $\kappa$ and $\Gamma$ are indicated). Symbols with error bars correspond to the results from MD simulations.~\cite{OhtaPRL2000,HamaguchiPS2001} Solid curves are calculated using Eq.~(\ref{T2}). Relaxation time is estimated from Eq.~(\ref{TR2}).}
\label{Fig2}
\end{figure}

As a next example we consider transverse mode dispersion data obtained by Ohta and Hanmaguchi.~\cite{OhtaPRL2000,HamaguchiPS2001}
They performed MD simulations of collective modes in strongly coupled Yukawa fluids, in the vicinity of the fluid-solid phase transition (see e.g. Fig.~4 from Ref.~\onlinecite{KhrapakPRE2014}).
The comparison between these numerical results and the analytical dispersion of Eq.~(\ref{T2}) is shown in Fig.~\ref{Fig2}. The agreement is very good, especially in the range $q\lesssim 2$.

\begin{figure}
\includegraphics[width=85mm]{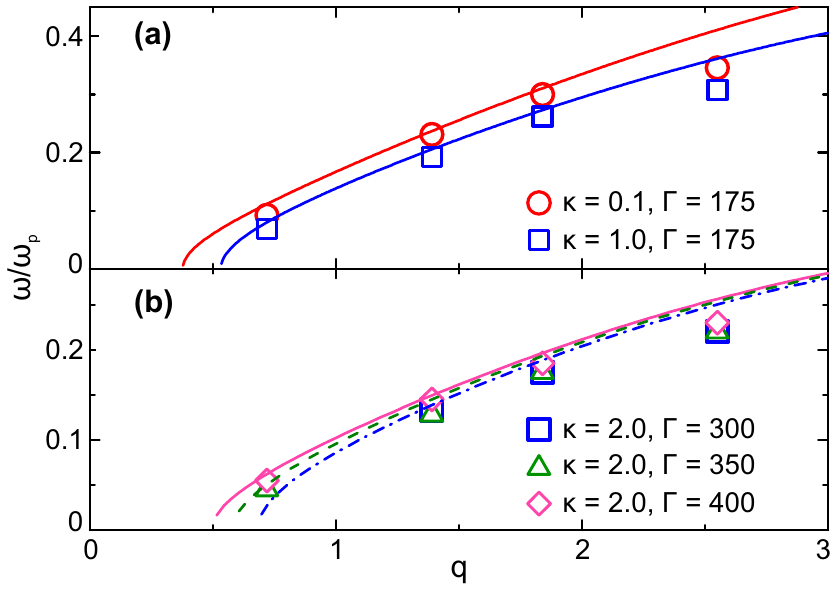}
\caption{Transverse mode dispersion of strongly coupled Yukawa fluids for the following set of $\kappa$ and $\Gamma$ values: (a) $\kappa=0.1$ (circles) and $1.0$ (squares) at a fixed coupling parameter $\Gamma= 175$; (b) $\Gamma = 300$ (squares), 350 (triangles), and 400 (rhombs) at a fixed screened parameter $\kappa = 2.0$. Symbols correspond to numerical data.~\cite{MithenPRE2014} Curves of the corresponding color are plotted using Eqs.~(\ref{T2}) and (\ref{TR2}).
}
\label{Fig3}
\end{figure}

In Fig.~\ref{Fig3} we show the comparison between shear waves dispersion of Yukawa fluids, reconstructed from the transverse current correlation function using MD simulations by Mithen,~\cite{MithenPRE2014} for five strongly coupled state points.  The agreement between the numerical and theoretical results is again convincing for $q\lesssim 2$.

Note that in the regime where the agreement is particularly good ($q\lesssim 2$) the deviations from the acoustic dispersion are not particularly important. This suggests that a simple expression (\ref{Hyd}) can also be used in the zero approximation. Another related point is that the main source of discrepancy between the theoretical and numerical data at $q\gtrsim 2$ is related to some worsening of the QCA approximation itself, not to the neglect of $q$-dependence of $\tau_{\rm R}$. The location of the first Brillouin pseudo-zone boundary can be estimated from the condition $(4\pi/3)(q_{\rm B}/2\pi a)^3=n$, which results in $q_{\rm B}=(9\pi/2)^{1/3}\simeq 2.42.$

One more opportunity to check the relevance of our approach is to analyze the behavior of the cutoff wave number $q_*$. Quite generally, the $q$-gap widens on moving away from the melting point (that is by increasing temperature or lowering $\Gamma$).
For Yukawa fluids, quantitative dependence for $q_*$ has been obtained by Goree {\it et al.} using MD simulations.~\cite{GoreePRE2012} They identified the wave number corresponding to the onset of a negative peak in the transverse current correlation function. In this way it was demonstrated that $q_*$ is a quasi-universal function of the properly normalized coupling parameter. An approximation of the form $q_*\simeq \tfrac{1}{3}(\Gamma/\Gamma_{\rm m})^{-4/3}$ was proposed.~\cite{GoreePRE2012}

\begin{figure}[!t]
\includegraphics[width=85mm]{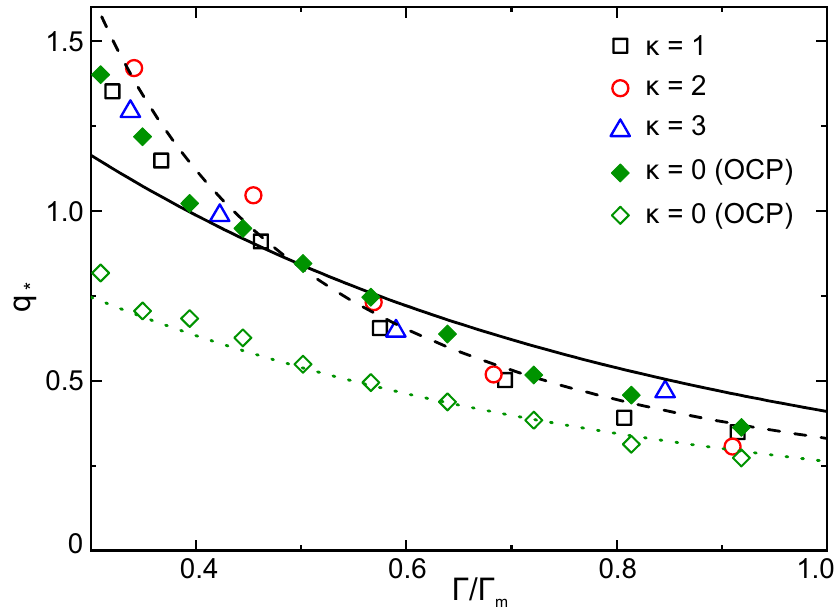}
\caption{Cutoff wave number $q_*$ for shear waves in strongly coupled Yukawa fluids versus the coupling parameter $\Gamma$, normalized to its value at the fluid-solid phase transition $\Gamma_{\rm m}$. Symbols correspond to MD simulation results~\cite{GoreePRE2012} for $\kappa=1$ (squares), 2 (circles), and 3 (triangles). Solid and open rhombs are our own MD results for the OCP fluid (for discussion see the text). The dashed line is the approximation proposed in Ref.~\onlinecite{GoreePRE2012}. Solid and dotted curves are plotted using the functional form of Eq.~(\ref{q_scal}) from the present article.}
\label{Fig4}
\end{figure}

If the cutoff wave number is sufficiently small (so that $\omega_{\rm QCA}\simeq c_t(q/a)$ and $q$-dependence of $\eta$ and $\tau_{\rm R}$ can be safely neglected), the cutoff wave number can be expressed as
\begin{equation}
q_*\simeq 0.77\frac{(c_t/\omega_{\rm p}a)\sqrt{\Gamma}}{\eta_*}.
\end{equation}
Sound velocities of strongly coupled Yukawa fluids (both longitudinal and transverse), when expressed in units of $\omega_{\rm p}a$ are known to be extremely weak functions of $\Gamma$ and only $\kappa$ dependence is important.~\cite{KalmanPRL2000,KhrapakPRE2015_Sound,KhrapakPPCF2016}  Moreover, we have found out that the product of $\kappa$-dependent quantities ($c_t/\omega_{\rm p}a$) and $\sqrt{\Gamma_{\rm m}}$ is practically constant, equal to $2.55\pm 0.10$ in the regime $\kappa\lesssim 5$. Combining this with Eq.~(\ref{visc}) for the reduced viscosity coefficient we arrive at the following scaling
\begin{equation}\label{q_scal}
q_*\simeq 15.6\sqrt{\Gamma/\Gamma_{\rm m}}\exp\left(-3.64\sqrt{\Gamma/\Gamma_{\rm m}}\right).
\end{equation}
The derived scaling holds for $q_*\lesssim 1$, that is for $\Gamma/\Gamma_{\rm m}\gtrsim 0.4$. Eq.~(\ref{q_scal}) is plotted in Fig.~\ref{Fig4} by the solid curve. In the considered regime it is in reasonable agreement with numerical results and approximate expression reported by Goree {\it et al.}~\cite{GoreePRE2012}

In Figure~\ref{Fig4} we also show our own MD results (see Appendix) for the cutoff wave number in the OCP limit ($\kappa=0$).  The solid rhombs correspond to peak positions of the transverse-current autocorreletion function obtained numerically. These results agree well with the previous results for Yukawa systems.~\cite{GoreePRE2012} The open rhombs have been obtained using a different procedure to be discussed below. This procedure yields somewhat lower threshold for the onset of transverse wave. Further details will be given in Sec.~\ref{Sqgap}.

Since most of the plasma-related specifics has been lost on arriving at Eq.~(\ref{q_scal}), it might be interesting to check the melting-temperature version of this scaling ($\Gamma/\Gamma_{\rm m}\rightarrow T_{\rm m}/T$) for conventional neutral fluids.
In particular, from the presented results it follows that the reduced cutoff wave-number at the melting point reaches a quasi-universal value of $q_*\simeq 0.3-0.4$. The relevance of this result beyond plasma-related (Yukawa) context will be discussed below.

\section{Application to liquid metals}

As an example of application of the approach to neutral fluids we consider the experimental data on transverse excitations in liquid Fe, Cu, and Zn recently reported by Hosokawa {\it et al}.~\cite{Hosokawa2015} Transverse acoustic (TA) modes have been detected experimentally using inelastic x-ray scattering through the quasi-TA branches in the longitudinal currant correlation spectra. In addition, the elastic constants and sound velocities have been estimated from experimental measurements.

\begin{table}[!b]
\caption{\label{Tab1} Properties of the liquid Fe, Cu, and Zn, relevant for the experiments by Hosokawa {\it et al}.~\cite{Hosokawa2015}  The melting temperature $T_{\rm m}$ and the shear viscosity coefficient at the melting temperature $\eta_{\rm m}$ are taken from Ref.~\onlinecite{Battezzati1989}. The elastic longitudinal and shear moduli ($M$ and $G$) along with the longitudinal and transverse sound velocities ($c_l$ and $c_t$) and the adiabatic sound velocity $c_s$ have been  tabulated in Ref.~\onlinecite{Hosokawa2015}. The instantaneous sound velocity $c_B=\sqrt{c_l^2-\tfrac{4}{3}c_t^2}$ is rather close to the adiabatic sound velocity.}
\begin{ruledtabular}
\begin{tabular}{lrrr}
Metal & Fe & Cu & Zn  \\ \hline
$T_{\rm m}$ (K) &  1809 & 1356  & 693  \\
$M$ (GPa) & 132 & 117 & 74    \\
$G$ (GPa) &  24 & 25 & 17  \\
$c_l$ (m/s) & 4330  & 3830 & 3360 \\
$c_t$ (m/s) & 1860  & 1790 & 1630 \\
$c_B$ (m/s) & 3760 & 3220 & 2780 \\
$c_s$ (m/s) & 3800 & 3460 & 2780 \\
$\eta_{\rm m}$ (mPa s) & 5.5 & 4.0 & 3.85\\
\end{tabular}
\end{ruledtabular}
\end{table}

\begin{figure}
\includegraphics[width=85mm]{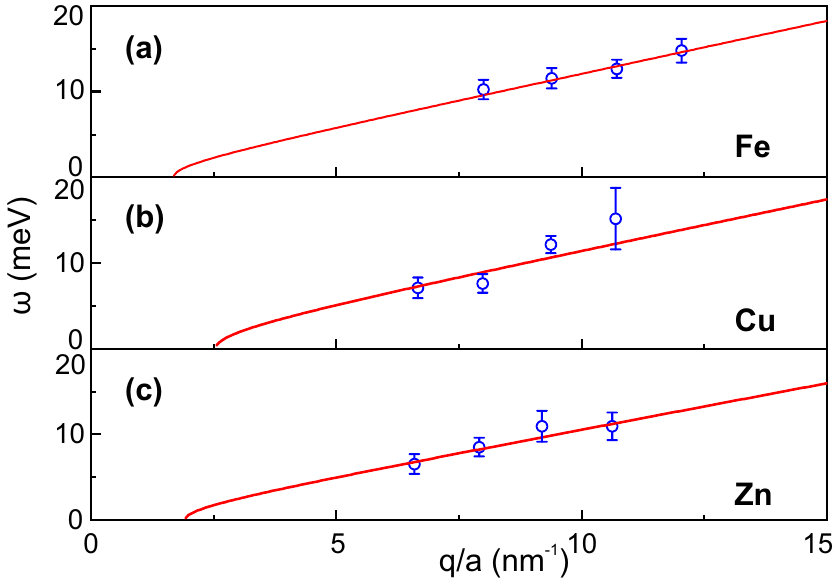}
\caption{The dispersion relation of the TA mode in liquid Fe (a), Cu (b), and Zn (c) near the melting temperature. The circles correspond to the experimental results by Hosokawa {\it et al}.~\cite{Hosokawa2015} The curves are plotted using Eqs.~(\ref{tauR}) and (\ref{Hyd}) of the present paper. The location of the first Brillouin pseudo-zone boundary is $q_{\rm B}/a \simeq 16.5$ nm$^{-1}$ (Fe and Cu) and $q_{\rm B}/a \simeq 15.3$ nm$^{-1}$ (Zn), beyond the range shown. }
\label{Fig5}
\end{figure}

The parameters corresponding to the experimental conditions of Ref.~\onlinecite{Hosokawa2015} are summarized in Table~\ref{Tab1}.
Since the experiments have been performed at temperatures just above the melting temperature, $T/T_{\rm m}=1.01$ (Fe), 1.05 (Cu), and 1.03 (Zn), the experimental value of the shear viscosity coefficient at the melting temperature can be used in calculations in the first approximation (note that in a recent study the viscosity of supercooled metallic melts has been related to the steepness of the short-range ion-ion repulsive interaction~\cite{KrausserPNAS2015}). The liquid densities are directly related to elastic moduli via $mn=M/c_l^2=G/c_t^2$, where $M$ and $G$ are the elastic longitudinal and shear moduli.
This yields the values close to those at the melting temperature.~\cite{Battezzati1989}
Taking these parameters, the long-wavelength portion of the dispersion relation has been calculated from Eq.~(\ref{Hyd}) and plotted in Fig.~\ref{Fig5}. Not unexpectedly, good agreement is observed.

\begin{table}[!b]
\caption{\label{Tab2} Dimensionless parameters characterizing liquid Fe, Cu, and Zn in conditions relevant to the experiments by Hosokawa {\it et al}.~\cite{Hosokawa2015} The thermal velocities are $v_{\rm T}=519$ m/s (Fe), $430$ m/s (Cu), and $300$ m/s (Zn).}
\begin{ruledtabular}
\begin{tabular}{lrrr}
Metal & Fe & Cu & Zn  \\ \hline
$c_l/v_{\rm T}$  & 8.3  & 8.9 & 11.2 \\
$c_t/v_{\rm T}$  & 3.6  & 4.2 & 5.4 \\
%$\tau_{\rm R}c_tn^{1/3}$ & 1.2 & 0.8 & 1.0 \\
$q_*$ & 0.3 & 0.4 & 0.3\\
\end{tabular}
\end{ruledtabular}
\end{table}

Some relevant dimensionless quantities, which might be of interest are summarized in Table~\ref{Tab2}.
In particular we observe that the longitudinal sound velocity reaches the value of approximately $\sim 10v_{\rm T}$ near the melting temperature. This is close to the melting temperature adiabatic sound velocity of Yukawa fluids in the moderate screening regime~\cite{KhrapakJCP2016} ($\kappa\sim 4$) as well as of repulsive inverse-power-law (IPL) spheres ($\phi\propto r^{-\alpha}$) with $\alpha\sim 10$.~\cite{KhrapakJCP2016,KhrapakSciRep2017} At the same time, this is somewhat smaller than the adiabatic sound velocity of the hard sphere fluid at freezing~\cite{KhrapakJCP2016,RosenfeldJPCM_HS} as well as the longitudinal elastic sound velocity of the IPL fluids.~\cite{KhrapakSciRep2017}

The TA velocity $c_t$ is about one half of the longitudinal sound velocity $c_l$.  This implies  that the generalized Cauchy relation is {\it not satisfied} in these experiments. The generalized Cauchy relation for spatially isotropic liquids with pairwise interactions reads~\cite{ZwanzigJCP1965,Schofield1966}
\begin{equation}
M-3G = 2(P-nT).
\end{equation}
On the scale of $M$ and $G$ values (see Tab.~\ref{Tab1}), the pressure term is negligible. The ideal gas term $nT$ is larger, $nT\sim {\mathcal O}(1 {\rm GPa})$ but still not very important. This implies $M\simeq 3G$ and, hence, $c_l\simeq \sqrt{3}c_t$.
The latter condition was previously used to estimate the transverse sound velocity of liquid cesium near the melting point.~\cite{MorkelJPCM1990,BodensteinerPRA1992,MorkelPRE1993} We observe, however, that it is violated in the considered case of Fe, Cu, and Zn near the melting temperature.

On the other hand, by analogy with elastic waves in solids,~\cite{LL} the high-frequency (instantaneous) bulk modulus can be introduced using $B=M-\tfrac{4}{3}G$. The (instantaneous) sound velocity defined with the help of the instantaneous bulk modulus, $c_B=\sqrt{B/nm}$, has been in many cases demonstrated to be very close to the conventional adiabatic sound velocity. This has been for instance documented for repulsive Yukawa and IPL systems near the fluid-solid phase transition.~\cite{KhrapakPoP2016_Relations,KhrapakSciRep2017,KhrapakPRE2018}  We observe that $c_ B\simeq c_s$ also for the considered liquid metals (see Tab.~\ref{Tab1}). This may therefore represent more conventional way of estimating the transverse sound velocity from the measured adiabatic and longitudinal velocities. For example, applying this approach to liquid cesium near melting we can obtain $c_t\simeq 500$ m/s ($c_l\simeq 1120$ m/s and $c_s\simeq 965$ m/s according to Refs.~\onlinecite{BodensteinerPRA1992,MorkelPRE1993}), whilst $c_l/\sqrt{3}$ amounts to $\simeq 650$ m/s.

The last row of Table~\ref{Tab2} indicates that the reduced critical wave-number at the onset of the transverse mode can be a quasi-universal quantity $q_*\simeq 0.3-0.4$ for various systems at the melting point. We remind that for Yukawa melts $q_*\simeq 0.4$ according to Eq.~(\ref{q_scal}). Combined with Eq.~(\ref{q*}) this implies that $\tau_{\rm R}\simeq 2 a/c_t$. This means that near the melting temperature the relaxation time is roughly the time needed for a shear wave to propagate a distance corresponding to one interparticle separation (diameter of the Wigner-Seitz cell). This observation might be quite useful in analyzing various properties of simple melts. There seems, however, considerable scattering present. For Yukawa melts we get $\tau_{\rm R}c_t/a\simeq 1.8$, while for liquid metals near the melting temperature
 $\tau_{\rm R}c_t/a \simeq 2.8$ (Fe), $1.9$ (Cu), and $2.4$ (Zn).

\section{Important remark on the cutoff wave number reconstruction}
\label{Sqgap}

%Equation \eqref{T2} is obtained from the generalized hydrodynamic approach and, thus, it is more suitable in the intermediate range of wavenumbers, where $\omega \gg \tau^{-1}$ and corresponding high-frequency collective excitations are enough well defined.
%This fact is also related to the propagation of high-frequency sound.
%However, at low frequencies of transverse excitations, that corresponds to the edge of band gap, the excitations are strongly damping and the results for $q$-gap become depending on the way in which it is calculated.
%However, at the temperatures near melting line, results obtained from different approaches are close with each other.

The real central problem for a proper calculation of $q$-gaps in fluids is that the transverse current autocorrelation function $C_{t}(q,\omega)$ should be properly analyzed in order to reconstruct correctly the dispersion curves $\omega(q)$. 
In case of most crystals with almost harmonic collective excitations,~\cite{YurchenkoJCP2018, YurchenkoPRE2017, KryuchkovPRL2018, HosokawaJPCM2013} this task is trivial:
$C_{t}(q,\omega)$ distributions represent narrow peaks, directly associated with the dispersion relation $\omega(q)$.
However, in case of fluids, the strong effect of anharmonicity results, in particular, in strong damping of collective excitations.
Due to this damping, the peak structure of $C_{t}(q,\omega)$ is lost and they become much broader.~\cite{YurchenkoPRE2017, KhrapakPRE2018} In this case, an appropriate theoretical model taking into account competition between  processes giving rise to mode propagation and damping is required.  

So far in this article, the discussed dispersion relations were mostly associated with the {\it maximum position} of $C_{t}(q,\omega)$ at a given $q$. This includes the generalized hydrodynamics approach, summarized in Sec.~\ref{Theo}, as well as the dispersion relations for the strongly coupled OCP and Yukawa fluids shown in Figs.~\ref{Fig1}, \ref{Fig2}, and \ref{Fig3} (the dispersion relations of TA modes in liquid metals shown in Fig.~\ref{Fig5} were obtained through the quasi-TA branches in the longitudinal current correlation spectra, see Ref.~\onlinecite{Hosokawa2015} for details). It appears that identification of dispersion curves using the maxima of $C_{t}(q,\omega)$ becomes problematic near the $q$-gap boundary (as well as when coupling weakens). It has been suggested previously that a more reasonable model to describe the transverse current spectrum is the damped harmonic oscillator (DHO) model.~\cite{HansenBook,FakPBCM1997, AliottaPRE2011} In this case the transverse current correlation function is proportional to the sum of two Lorentzian terms:
\begin{equation}\label{fit}
C_{t}(\omega)\propto\frac{1}{(\omega-\omega_{\rm T})^2+\gamma_{\rm T}^2}+\frac{1}{(\omega+\omega_{\rm T})^2+\gamma_{\rm T}^2}.
\end{equation}
As a result, fitting the MD generated $C_{t}(q,\omega)$ with the functional form of Eq.~(\ref{fit}) for each value of $q$ we obtain the dispersion curve $\omega_{\rm T}(q)$ and the damping rates $\gamma_{\rm T} (q)$ characterizing excitations' lifetime. Generally, the dispersion curve $\omega_{\rm T}(q)$ is not identical to $\omega_t(q)$ associated with the second frequency moment, although in certain regimes they are close to each other (and this is exactly what the QCA theory claims). Note that     
the maximum of $C_t(\omega)$ defined via Eq.~(\ref{fit}) occurs at $\omega = 0$ for $\omega_{\rm T} < \gamma_{\rm T}/\sqrt{3}$, even if $\omega_{\rm T}$ is non-zero itself, which corresponds to the overdamped  collective fluctuations. Typically, $\omega_{\rm T}$ and $\gamma_{\rm T}$ are of the same order in fluids,~\cite{KhrapakJCP2018, KhrapakPRE2018} especially far from the melting point (and at large $q$), which corresponds to the cases of (i) high temperature, (ii) low pressures, (iii) low densities. Because of this the dispersion relation evaluated from the maxima of $C_{t}(\omega)$ can appear somewhat lower than $\omega_{\rm T}$ and reaches zero value faster, resulting in larger values of the cutoff wave-number $q_*$.

\begin{figure}[!t]
\includegraphics[width=85mm]{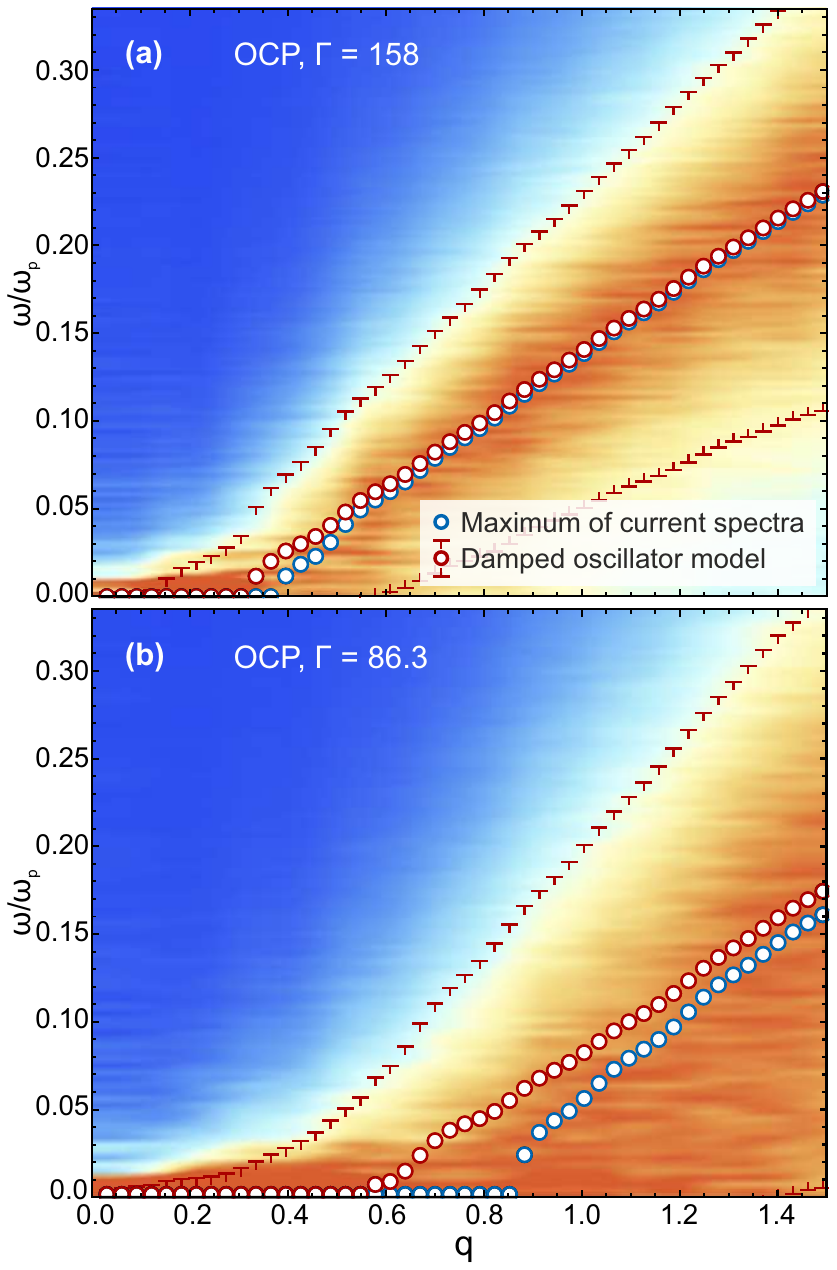}
\caption{Dispersion relation of the transverse mode of the strongly coupled OCP fluid with $\Gamma= 158$ (a) and $\Gamma=86.3$ (b). Blue and red circles are the frequencies $\omega_{\rm T}$ obtained from our MD simulations using the maxima of the transverse current spectra and the DHO model, respectively. Bars indicate the range $(\omega_{\rm T}\pm \gamma_{\rm T})$, where $\gamma_{\rm T}$ is the damping rate.
}
\label{Fig6}
\end{figure}

This is illustrated in Fig.~\ref{Fig6}, where the dispersion relations of the transverse mode in the strongly coupled OCP fluid are plotted for the two values of the coupling parameter, $\Gamma = 158$ (a) and $\Gamma = 86.3$ (b). The details of our MD simulations are summarized in the Appendix. We observe that the difference in determining the dispersion relation from the maxima  of $C_{t}(\omega)$ and the DHO model are rather small at strong coupling, but nevertheless visible near the onset of the transverse mode ($q$-gap boundary). At weaker coupling, the difference becomes much more pronounced. 

Further results from our MD simulation of the transverse waves in the OCP fluid are summarized in Fig.~\ref{Fig4}. Cutoff wave-numbers $q_*$ measured using the maxima of  $C_{t}(\omega)$ are depicted by solid rhombs. They are grouped quite close to the universal scaling with melting temperature (dashed curve) reported for Yukawa fluids in Ref.~\onlinecite{GoreePRE2012}. The black solid curve, corresponding to Eq.~(\ref{q_scal}), reproduces fairly well the near-melting behavior, but becomes inappropriate at lower $\Gamma/\Gamma_{\rm m}$, as expected. The cutoff wave-numbers $q_*$, obtained using the DHO model are depicted by the open rhombs in Fig.~\ref{Fig4}. They are located below all the other symbols. Interestingly enough, however, the temperature dependence of $q_*$ can still be well reproduced by Eq.~(\ref{q_scal}) with a re-scaled ($\simeq 10$ instead of $15.6$) front factor (shown by dotted curve). 

Detailed analysis of the $q$-gap properties in different types of fluids and in different thermodynamic regimes is an important separate problem, which is, however, beyond the scope of the present article.

\section{Conclusion}

The main results from this study are as follows. We have presented a simple practical approach to describe the dispersion of the transverse mode in strongly coupled plasma fluids. The approach is especially designed to situations when the concept of one-component Yukawa fluid is relevant. It works well at long wavelengths and properly accounts for the existence of the critical wave-number below which the transverse mode is absent ($q$-gap).  Extensive comparison with numerical simulations demonstrates that the accuracy of the approach should be sufficient for most practical purposes.

Although developed mostly for Yukawa fluids, the approach can be also of some use in other circumstances. Towards the end of the article we have illustrated this by considering the existing experimental data on transverse excitations in some liquid metals near the melting temperature. Some related scalings and emerging quasi-universalities have been discussed.

Finally, we pointed out to the importance of using an appropriate model (fitting function) for strongly anharmonic collective modes in fluids. In particular, we demonstrated that transverse dispersion relations obtained using the damped harmonic oscillator model can differ considerably from the traditional ones obtained from the maxima of transverse current correlation spectra. The difference, being relatively small near the fluid-solid phase tradition, becomes more and more pronounced away from the melting point. In this way the actual location of the $q$-gap boundary for transverse waves propagation may be overestimated.        

Among the potential applications of the approach we can mention the following.
Shear waves can be excited, observed, and analyzed in various dusty plasma experiments.~\cite{PramanikPRL2002,PielPoP2006,BandyopadhyayPLA2008} The cutoff wave number of shear wave was also measured experimentally (although in a 2D dusty plasma fluid).~\cite{NosenkoPRL2006} Here the approach can serve as a useful link between the observed quantities and the dusty plasma parameters (i.e. for diagnostic purposes). Another important aspect deals with the relations between the collective modes and transport properties of simple liquids.~\cite{ZwanzigJCP1983} Here an appropriate model for the transverse dispersion relation accounting for the $q$-gap may be essential in constructing convincing quantitative models. The same obviously applies to recent attempts to develop the phonon theory of liquid thermodynamics.~\cite{BolmatovSciRep2012,Trachenko2015}

\begin{acknowledgments}
We thank Viktoria Yaroshenko for careful reading of the manuscript.
MD simulations at BMSTU were supported by the Russian Science Foundation, Grant No. 17-19-01691. AGK acknowledges support  from the presidium RAS within the framework of the program No. 13 ``Condensed Matter and Plasma at High Energy Densities''. 
\end{acknowledgments}

\appendix*
\section{MD simulations}
\label{A_MD}
We have performed MD simulation of the OCP fluid ($\lambda\rightarrow\infty$ limit of the Yukawa interaction potential) in $NVT$ ensemble to measure the properties of transverse modes and, in particular,  the dependence of the cutoff wave number $q_\ast$ on the coupling parameter $\Gamma$. We have considered three-dimensional fluid consisting of $N=10^4$ particles in a cubic domain with periodic boundary conditions. To account for the long-range Coulomb interaction, the PPPM approach~\cite{LeBardSoftMatter2012} with the cut-off radius of $7.5n^{-1/3}$ for the short-range part has been employed. The numerical time step has been chosen as $\Delta t=5.1\times10^{-4}\sqrt{ma^3\Gamma/Q^2}$. All simulations have been run for $2\times10^6$ time steps, where the first half of the simulation time is used to equilibrate the system and the second one is used to calculate excitation spectra.
Simulations have been performed using HOOMD-blue package.~\cite{AndersonJCompP2008, GlaserJCompP2015}

Excitation spectra in the fluid phase have been obtained based on the standard approach, employed previously in Refs.~\onlinecite{YurchenkoPRE2017, YurchenkoJCP2018, KryuchkovPRL2018}.
The transverse ($t$) current autocorrelation function $C_{t}(\mathbf{q},\omega)$ has been calculated from:
\begin{equation}
C_{t}(\mathbf{q},\omega)\propto\mathrm{Re}\int dt \left<j_{t}(\mathbf{q},t)j_{t}(-\mathbf{q},0)\right>e^{i\omega t},
\end{equation}
where  $j_t(\mathbf{q},t)$ is the projection of the particle current $\mathbf{j}(\mathbf{q},t)\propto\sum_s\mathbf{v}_s(t)\exp\left(i\mathbf{q}\mathbf{r}_s(t) \right)$ to the transverse direction, $\mathbf{v}_s(t)=\dot{\mathbf{r}}_s(t)$ is the velocity of $s$-th particle at a time $t$.  Due to isotropy of simple fluids, $C_{t}(\mathbf{q},\omega)$ corresponding to different directions of $\mathbf{q}$ can be averaged to suppress thermal noise.

\bibliography{KhrapakJan2019_References}

\end{document}